\documentclass[epjST]{svjour}
\usepackage{graphics}
\usepackage{amssymb}
\begin{document}
\title{A Dyson-Schwinger model beyond isospin limit }
\subtitle{prepared for investigating $U_A(1)$-breaking temperature dependence}
\author{Davor Horvati\'c\inst{1}\fnmsep\thanks{\email{davorh@phy.hr}} \and
 Dalibor Kekez\inst{2}\fnmsep\thanks{\email{kekez@irb.hr}}
 \and Dubravko Klabu\v{c}ar\inst{1}\fnmsep\thanks{\email{klabucar@phy.hr}}}
\institute{Physics Department, Faculty of Science-PMF, University of Zagreb, Bijeni\v{c}ka cesta 32,
10000 Zagreb, Croatia \and Rugjer Bo\v{s}kovi\'c Institute, Bijeni\v{c}ka cesta 54, 10000 Zagreb, Croatia}
\abstract{Motivated by our earlier findings of sensitive quark-flavor dependence
 of QCD topological susceptibility on products of current quark masses and
 corresponding condensates, we allow the breaking of isospin symmetry.
 For the purpose of future investigations of $U_A(1)$ symmetry breaking and
 restoration at $T > 0$, we perform (at $T=0$) refitting of the quark-mass
 parameters of a phenomenologically successful effective model of low-energy
 QCD. It belongs to the class of separable-interaction models within the
 Dyson-Schwinger approach to the quark-antiquark substructure of mesons.  
} %end of abstract
\maketitle
\section{Introduction and survey}
\label{intro}

In the realm of nonperturbative strong interactions, {\it ab initio} calculations
 are often of prohibitive difficulty. The usage of simplified models mimicking
 the underlying fundamental theory of QCD, is still often unavoidable in its
 low-energy, nonperturbative regime. This holds especially in applications at
 high temperature around or above the (pseudo-)critical temperature
 ($T \gtrsim T_{crit}$)
 and/or density, such as heavy-ion collisions, as well as in astrophysical and
 cosmological applications. The more convoluted a context of applications
happens to be, the stronger simplifications in modeling of dynamics are needed
in the quest for tractability, as long as some crucial properties of the underlying
 QCD are reproduced.
 Obviously, favored are models which are as simple as possible while achieving
 as much as possible. The case in point is the Dyson-Schwinger (DS) 
 separable-interaction model of Blaschke {\it et al.} \cite{Blaschke:2000gd}.

On the one hand, it makes predictions for so much low-energy phenomenology
that it is competitive, in the low-energy and momentum regime, with some
more elaborate model interactions (incorporating also the high-energy part,
see {\it corresponding} predictions in, {\it e.g.}, \cite{Klabucar:1997zi,Kekez:1998rw,Kekez:2000aw,Kekez:2003ri,Kekez:2005ie}),
while on the other hand, it reproduces the proper chiral behavior of QCD. The
 latter quality, shared with other consistently applied DS approaches to QCD
but very rare or even absent in other quark-bound-state approaches to hadrons,
is usually even more important than the former. It is certainly so when
 dealing with
light pseudoscalar mesons and issues concerning the dynamically broken
$SU_A(N_f)$ chiral symmetry and the related $U_A(1)$ symmetry and anomaly.

Thus, chirally correct DS models exhibiting (at high $T$, but at vanishing
and low density) a second order transition in the chiral limit and a crossover
for realistic $u,d,s$-quark masses (i.e., $N_f = 2+1$) are well suited for
modeling of low-energy QCD. The presently used model, stemming from 
Blaschke {\it et al.} \cite{Blaschke:2000gd}, served very well for the
task of extending our DS-approach studies of $U_A(1)$ breaking through
$\eta'$-$\eta$ complex at $T=0$ \cite{Klabucar:1997zi,Kekez:2000aw,Kekez:2003ri,Kekez:2005ie,Horvatic:2007mi,Benic:2014mha}
 to nonvanishing temperatures \cite{Horvatic:2007wu,Horvatic:2007qs,Horvatic:2010md,Benic:2011fv,Horvatic:2018ztu,Horvatic:2019eok}.
In the next section, we present the results of the re-fitting,
out of the isospin-symmetric limit, of the quark mass parameters 
in the model's so-called rank-2 variant, used in \cite{Horvatic:2007wu,Horvatic:2007qs,Horvatic:2010md,Benic:2011fv,Horvatic:2018ztu,Horvatic:2019eok}.

Of course, the isospin symmetry holds very well for almost all purposes
 in the context of hadronic physics. Let us, however, observe that this
 led, in general, to a somewhat cavalier attitude among hadronic model
users, where the usual traditional practice has been to fit the experimental
charged pion and kaon masses ${M}_{\pi^\pm,K^\pm}^{\rm exp}$ in models
 mimicking only QCD,
 even though the QED contributions can be taken as vanishing only for
 neutral pions and kaons due to Dashen's theorem \cite{Dashen:1969eg},
 which is not violated much \cite{Aoki:2013ldr}. Then, the 
 electromagnetic contribution would be there only for charged pions
(around 4.5 -- 4.7 MeV  \cite{Aoki:2013ldr,Gasser:1982ap,Donoghue:1996zn})
 and charged kaons (around 1.3 -- 2.5 MeV
 \cite{Aoki:2013ldr,Gasser:1982ap,Donoghue:1996zn}).
% charged pions, and around 1.3 -- 2.5 MeV for charged kaons, see,
% {\it e.g.}, Refs. \cite{Gasser:1982ap,Donoghue:1996zn,Aoki:2013ldr}.
Hence, if a model only mimics QCD, one should fit it to hadron
masses out of which the electromagnetic contribution has been taken out.
(Let us distinguish such masses by carets: $\hat{M}_{\pi^\pm}$,
$\hat{M}_{K^\pm}$, etc.)

 If one abandons the isospin symmetry, one should find the three
different light quark masses $m_u$, $m_d$ and $m_s$ by fitting the
three meson masses. These are the masses (\ref{EMsubtractedM}) of
% the charged pion, charged kaon
$\pi^+$, $K^+$ and the neutral kaon $K^0$, {\it out of which the
electromagnetic contributions were taken out}, because the model
is of QCD only. The appropriate results are given by the FLAG
collaboration \cite{Aoki:2013ldr} in their Eq. (11) for the pion
and kaon masses occurring in just the QCD sector of the Standard
Model \cite{Aoki:2013ldr}, {\it i.e.}, with QED turned off:
\begin{equation}
\hat{M}_{\pi^+}= 134.8 \pm 0.3\, \mbox{MeV}~, 
\quad \hat{M}_{K^+}= 491.2 \pm 0.5\,\mbox{MeV}~, 
\quad \hat{M}_{K^0}=497.2 \pm 0.4\,\mbox{MeV}~.
\label{EMsubtractedM}
\end{equation}
By contrast, the traditional fit would be to the corresponding
 experimental values (rounded to one decimal place):
\begin{equation}
{M}_{\pi^+}^{\rm exp}= 139.6\,\, \mbox{MeV}~, 
\qquad {M}_{K^+}^{\rm exp} =  493.7\,\, \mbox{MeV}~, 
\qquad {M}_{K^0}^{\rm exp} =497.6 \,\,\mbox{MeV}~.
\label{experMase}
\end{equation}
For the both cases of refitting, in the next section we predict the
concrete model mass of $\pi^0$, along with the masses of the other
 two flavorless pseudoscalars $\eta'$ and $\eta$, after we
 incorporate the anomalous $U_A(1)$ breaking. 

Concerning the concrete presently used model, the reference
\cite{Blaschke:2000gd} had pertained only to the non-strange sector,
and upon including $s$-quarks starting with Ref. \cite{Horvatic:2007wu}
some re-fitting was already done. The model details including the
parameter values we were using earlier \cite{Horvatic:2007wu,Horvatic:2007qs,Horvatic:2010md,Benic:2011fv,Horvatic:2018ztu,Horvatic:2019eok},
are listed in the Appendix of our Ref. \cite{Horvatic:2019eok}. 
Everything so far has been in the isospin limit of equal $u$- and $d$-quark masses,
which is usually completely adequate in hadronic physics. However,
extending to $T>0$ our treatment the $U_A(1)$ anomaly contribution (needed
for $T$-dependence of $\eta'$ and $\eta$ mesons \cite{Horvatic:2018ztu}
and of axions \cite{Horvatic:2019eok}), led us to expressions involving the
harmonic average of the products of light quark masses and condensates:
the light-quark expression for QCD topological susceptibility and the
anomalous contribution of the masses in the $\eta'$-$\eta$ complex
(respectively, Eq. (19) and Eq. (18) in Ref. \cite{Horvatic:2018ztu},
as well as (\ref{chiShore_small_m}) and (\ref{defA}) below).
The harmonic average is dominated by the lightest flavor to such an
extent that in Ref. \cite{Horvatic:2018ztu} we concluded one should
check whether the isospin breaking between $u$- and $d$-quark masses
can significantly affect the $T$-dependence of the $U_A(1)$ anomaly
mass contribution.
This, and not a search for a better description of the light pseudoscalar
nonet masses, is the reason that as the first step (at $T=0$) we perform
the re-fitting of the quark mass parameters of our Dyson-Schwinger model
of choice, but without the constraint of the  isospin symmetry.

The model calculation and its results at $T=0$ are  presented in the next section.

\section{The model calculation and the present re-fitting}
\label{results}

All model details, except of course the quark-mass parameter values,
can be found in one place -- in the Appendix of our Ref. \cite{Horvatic:2019eok}.
From there, we adopt as the interaction model the effective gluon propagator
in a Feynman-like gauge and in the separable form:
\begin{equation}
g^2 \, D_{\mu\nu}^{ab}(p-\ell)_{\mbox{\rm\scriptsize eff}} \, = \,
\delta^{ab} \, g^2 \, D_{\mu\nu}^{\mathrm{eff}} (p-\ell) \, \longrightarrow \,
\delta_{\mu\nu} \, D(p^2, \ell^2,p\cdot \ell) \, \delta^{ab} \, .
\label{Feynmangauge}
\end{equation}
We choose the separable interaction variant called rank-2, 
\begin{equation}
D(p^2,\ell^2,p\cdot \ell) \, = \, D_0 \, {\cal F}_0(p^2) \, {\cal F}_0(\ell^2)
\, + \, D_1 \, {\cal F}_1(p^2) \, (p\cdot \ell ) \, {\cal F}_1(\ell^2)~,
\label{sepAnsatz}
\end{equation}
because it was used also in Ref. \cite{Horvatic:2018ztu} on $T$-dependence of
$\eta'$ and $\eta$ mesons. The momentum-dependent functions ${\cal F}_0(p^2)$
and ${\cal F}_1(p^2) \, $ \cite{Horvatic:2007wu,Blaschke:2007ce} are
\begin{equation}
%\label{F0}
{\cal F}_0(p^2) = \exp(-p^2/\Lambda_0^2)
\qquad {\rm and} \qquad
{\cal F}_1(p^2) = \frac{1+\exp(-p_0^2/\Lambda_1^2)}
{1+\exp((p^2-p_0^2)/\Lambda_1^2)}~,
%\label{F1}
\label{F0and1}
\end{equation}
where $D_0\Lambda_0^2 = 219$, $D_1\Lambda_0^4 = 40$,
$\Lambda_0=0.758$~{GeV}, $\Lambda_1=0.961$~GeV and $p_0=0.6$~{GeV}.
These values of the parameters of the interaction in the present work
are the same as in Refs. \cite{Horvatic:2018ztu,Horvatic:2019eok}, 
because in the present paper we vary only the quark mass parameters $m_q$
$(q = u,d,s)$ away from the isosymmetric values $m_u=m_d \equiv m_l=5.49$~MeV
and $m_s=115$~MeV, used in Refs. \cite{Horvatic:2018ztu,Horvatic:2019eok}.

All calculations are done as in Ref. \cite{Horvatic:2018ztu} up to
the point where the isospin symmetry leads to simplifications due to
$m_u = m_d$, from which we now refrain and take $m_u \neq m_d$.

Re-fitting is an arduous procedure, but it is in principle
straightforward to vary values of $m_q$'s into the gap equations for
dressed quark propagators of different flavors and then, in turn, into
the consistent Bethe-Salpeter equations for $q'\bar q\,$ $(q,q' = u,d,s)$
pseudoscalar bound-state vertices and masses $M_{q'\bar q}$. Through the
solutions of these DS equations, varying values of $m_q$'s affect all
calculated quantities calculated in Ref. \cite{Horvatic:2018ztu}, notably
in the condensates $\langle {\bar q} q\rangle$, which now all differ for
different flavors $q$ ({\it all} are flavor-nonuniversal now, while
before, $\langle {\bar u} u\rangle = \langle {\bar d} d\rangle$
  \cite{Horvatic:2018ztu,Horvatic:2019eok}).

\begin{figure}
\resizebox{0.75\columnwidth}{!}{%
  \includegraphics{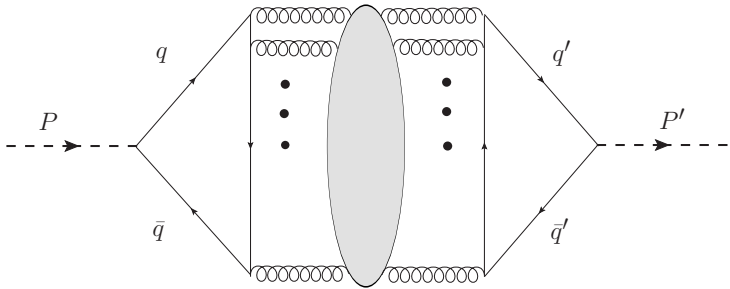} }
\caption{Flavor-changing, axial anomaly-driven  transitions of
quark-antiquark pseudoscalars $\, P=q\bar q\,$ to $\, P'=q'\bar q'\,$
comprising the pseudoscalar mesons in the hidden-flavor sector. All
quark and gluon lines and vertices are dressed nonperturbatively.
The gray oval and three dots stand for infinity of all intermediate
gluon states enabling such transitions. (The number of gluons
% symbolized by springs
must be an {\bf even} \cite{Kekez:2000aw} number; the simplest case
 is when this figure reduces to the ``diamond graph'', with no oval
 blob and just two gluons, albeit dressed nonperturbatively.)}
\label{fig:1}       % Give a unique label
\end{figure}

We obtain the chiral-limit-vanishing bound-state masses $M_{q'\bar q}$
 $(q',q=u,d,s)$ by solving
 consistent DS gap and Bethe-Salpeter equations in the rainbow-ladder
 approximation, and this cannot capture effects of the $U_A(1)$ anomaly.
 But in the flavorless, or hidden-flavor sector (where $q' = q$), the
 non-anomalous masses $M_{q'\bar q}$ cannot provide the whole story on
 the masses of flavorless pseudoscalars, since the $U_A(1)$ anomaly
 contributes through the flavor-changing transitions
 $|q\bar q\rangle \to |q' \bar q' \rangle $,
 like those depicted schematically in Fig. \ref{fig:1}.
 The famous example of the relatively very heavy $\eta'$ meson shows
 it is essential to include the anomalous $U_A(1)$ symmetry breaking
 at least at the level of the masses. We do it as described in Refs.
\cite{Klabucar:1997zi,Kekez:2000aw,Kekez:2005ie,Horvatic:2007mi,Benic:2014mha},
{\it i.e.}, relying on the $U_A(1)$ anomaly being suppressed in the limit
of large number of QCD colors $N_c$ \cite{Witten:1979vv,Veneziano:1979ec}.
Thanks to this, the anomaly contribution ${\bf M}^2_{A}$  to the {\it total}
mass matrix (squared) ${\bf M}^2$ of the hidden-flavor complex
 $\eta'$-$\eta$-$\pi^0$, can be treated formally as a perturbation.
 In the lowest order, it is simply added
 \cite{Klabucar:1997zi,Kekez:2000aw,Kekez:2005ie}
 to the non-anomalous mass matrix (squared), ${\bf M}^2_{NA}$,
 made of the $M_{q'\bar q}$ contributions:
 $\, {\bf M}^2 =  {\bf M}^2_{NA} + {\bf M}^2_{A}~. $

The non-anomalous part of the mass matrix (in the basis $|q\bar{q}\rangle$)
is still \\
$\,\, {\bf M}^2_{NA} ={\bf diag} [M_{u\bar{u}}^2, M_{d\bar{d}}^2, M_{s\bar{s}}^2]~,\quad$
but now
 $M_{u\bar{u}} \neq M_{d\bar{d}} \neq M_{u\bar{d}} = M_{\pi^\pm}$.

What is nevertheless qualitatively different, is that since the
 isospin symmetry is not enforced, it no longer precludes a
 contribution to the neutral pion due to the $U_A(1)$ anomaly.
 It will contribute, albeit quantitatively very little,
 to $M_{\pi^0}$ and to the mass difference between $\pi^0$
 and $\pi^\pm$, so that $M_{\pi^0}^2$ will not be exactly
 ${1\over 2} (M_{u\bar{u}}^2 + M_{d\bar{d}}^2)$.
 This is because ${\bf M}^2_{A}$, the anomalous part of the mass
 matrix (squared) cannot any longer be reduced to the $2 \times 2$
 matrix of the isoscalar subspace of the $\eta'$-$\eta$ complex.

In the hidden-flavor sector, the flavor-changing transitions due to
 the $U_A(1)$ anomaly, $|q\bar q\rangle \to |q' \bar q' \rangle $,
 depicted schematically in Fig. \ref{fig:1},
 yield the matrix elements of the anomalous mass matrix (squared):
 \begin{equation}
\langle q\bar q | {\bf M}^2_A |q' \bar q' \rangle =
b_q \, b_{q'} \, ,
\label{elementM2AqqX}
\end{equation}
where $b_q \equiv \sqrt{\beta}$ for $q = u$. But, the amplitudes for the
transitions from, and into, lightest $u\bar u$ pairs are larger than
those for the significantly more massive $s\bar s$.  Thus, as in
our earlier papers, we allow for the effects of the breaking of the
SU(3) flavor symmetry for $q,q' = s$ by $b_s = X \sqrt{\beta}$, where
$X = f_{u\bar u}/f_{s\bar s}$ \cite{Horvatic:2018ztu}. However, now
we do it also for $q,q' = d$; namely, $b_d = Y \sqrt{\beta}$ and
 $Y = f_{u\bar u}/f_{d\bar d}$, even though it is clear that here
 in the mass matrix at $T=0$, the effect of the isospin breaking
 is small, due to $f_{u\bar u} \approx f_{d\bar d} \approx f_{\pi}$.
(Actually, here we define 
 $f_{\pi} = f_{\pi^0} = {1\over 2}(f_{u\bar u} + f_{d\bar d})$,
 since here we consider the anomaly only on the level of the masses
 and neglect its possible influence on meson decay constants.)

The total mass matrix of the hidden-flavor sector
  in the flavor basis $|q\bar q\rangle$ is
\begin{equation}
{\bf M}^2 \, = \, {\bf M}^2_{NA} \, + \, {\bf M}^2_A \, = \, 
\left[
\begin{array}{ccc}
 M_{{u\bar u}}^2 + \beta & \beta  Y &
   \beta  X \\
\beta  Y & M_{{d \bar d}}^2+\beta  Y^2
   & \beta  X Y \\
 \beta  X & \beta  X Y &
   M_{{s\bar s}}^2+\beta  X^2 \\
\end{array}
\right]
~,
\label{M2qqbarBasis}
\end{equation}
where, as in Refs. \cite{Benic:2014mha,Horvatic:2018ztu},
\begin{equation}
\label{defA}
\beta \, = \, \frac{2 A}{f_{\pi}^2}  \qquad  {\rm  and}  \qquad 
A \, = \, \frac{\chi}{\,\, 1 \, +\, {\chi} \, (\,\frac{1}{m_{u}\,\langle{\bar u}u\rangle}
 + \frac{1}{m_{d} \,\langle {\bar d}d \rangle}
+ \frac{1}{m_{s}\,\langle {\bar s}s \rangle } \,) \,} \, ,
\end{equation}
in line with Shore's generalization \cite{Shore:2006mm} of Witten-Veneziano
 relation \cite{Witten:1979vv,Veneziano:1979ec}, where $A$ is the full QCD
 topological charge parameter and where the QCD topological susceptibility
 $\chi$ is for light flavors given by the current masses $m_q$ multiplied
 by respective condensates $\langle {\bar q}q \rangle$ realistically away
 from the chiral limit (which gives the crossover behavior at large $T$):
\begin{equation}
\chi \, = \, \frac{- \, 1}{\,\,\,  \frac{1}{\, m_{u} \, \langle {\bar u}u \rangle } +
\frac{1}{m_{d}\,\langle{\bar d}d \rangle} + \frac{1}{m_{s}\,\langle {\bar s}s \rangle }
            \,\,\,  } \, + \, {\cal C}_m \, .
\label{chiShore_small_m}
\end{equation}
As before \cite{Benic:2011fv,Benic:2014mha,Horvatic:2018ztu}, the
small-magnitude and necessarily negative correction term ${\cal C}_m$
(higher order in small quark masses) is found by using (as well as
 Shore \cite{Shore:2006mm}) the $1/N_c$ approximation
 $A = \chi_{\mbox{\rm\scriptsize YM}}$ valid at $T=0$.
 As before, we adopt the lattice result 
$\chi_{\mbox{\rm\scriptsize YM}} = (191\, {\rm MeV})^4$ \cite{DelDebbio:2004ns}.

With these ingredients, the pseudoscalar meson states and masses
containing the influence of the $U_A(1)$ anomaly are readily obtained
by diagonalization of the complete mass matrix of the hidden-flavor
sector, Eq. (\ref{M2qqbarBasis}). The eigenvalues of this matrix
are the present model predictions for the squared physical masses
 of $\eta'$, $\eta$, and $\pi^0$, given in Table \ref{3M+2f}
 in the next subsection.

\subsection{Results out of the isospin symmetry limit}
\label{sec:2}

The values of the model quark mass parameters $m_q$ ($q=u,d,s$)
are obtained by fitting mass eigenvalues of Bethe-Salpeter equations
to various values $M_{\pi^\pm}^{\rm fit}$, $M_{K^\pm}^{\rm fit}$ and
 $M_{K^0}^{\rm fit}$, assigned to $\pi^\pm$, $K^\pm$ and $K^0$
 meson masses in three different ways described below.

In each of the tables presenting our model results, 
the first row corresponds to the isosymmetric case $m_u = m_d$.
It is here for comparison because it just repeats what
we already had (at $T=0$) in our previous references
\cite{Horvatic:2007mi,Benic:2014mha,Horvatic:2007wu,Horvatic:2007qs,Horvatic:2010md,Benic:2011fv,Horvatic:2018ztu,Horvatic:2019eok}
employing this model, where the parameters
 (including $m_u = m_d = 5.49$ MeV and $m_s = 115.12$ MeV)
were obtained through the fit of the pion and kaon masses in the isospin
limit, so that all pion masses, including $M_{\pi^\pm}^{\rm fit}$, were
taken equal (140 MeV), and all kaon masses were taken equal (495 MeV),
including $M_{K^\pm}^{\rm fit}$ and $M_{K^0}^{\rm fit}$.

The next two rows in each of the tables correspond to the results
of two different fits out of the isospin limit,
% both allowing 
$m_u \neq m_d$, but fitting somewhat differently defined pion
and kaon masses. The second row in every table is labeled by 
the superscript ${}^{(2)}$, because this corresponds to fitting
the values in Eq. (\ref{experMase}). This is the traditional fit
to the experimental values (here, rounded to the first decimal place).
It yields $m_u = 4.37$ MeV, $m_d = 6.55$ MeV and $m_s = 115.34$ MeV.


\begin{table}
\caption{The first three rows of numbers represent our three fits. The
 first block of columns are the masses to which the model quark-mass
 parameters $m_q$ were fitted. Next are the predicted observables:
 charged pion and kaon decay constants ($f_{\pi^+} = f_{u\bar d}$
 and $f_{K^+} = f_{u\bar s}$, respectively), and the masses of the
 flavorless pseudoscalars $\pi^0$, $\eta$ and $\eta'$. 
The last row gives the experimental values of all these quantities.
 All values are in MeV. }
\label{3M+2f}       % Give a unique label
\begin{tabular}{|c|ccc|cc|ccc|}
\hline
&&&&&&&&  \\
Type of fit& $M_{\pi^\pm}^{\rm fit}$ & $M_{K^\pm}^{\rm fit}$ & $M_{K^0}^{\rm fit}$ &  $f_{\pi^+}$ & $f_{K^+}$ & $M_{\pi^0}$ & $M_{\eta}$ & $M_{\eta'}$\\
&&&&&&&&  \\
\hline
&&&&&&&&  \\
$m_u=m_d$  & 140.0 &  495.0 &  495.0 & 92.0 & 108.8  & 140.0 &  554.0 & 997.0\\
&&&&&&&&  \\
 ${}^{(2)} m_u\! = 0.67 m_d $  & 139.6 &  493.7 &  497.6 & 92.0 & 108.5 & 139.6 &  554.5 & 995.5 \\
&&&&&&&&  \\
 ${}^{(1)} m_u\! = 0.5\, m_d $  & 134.9 &  492.6 &  498.7 & 91.8 & 108.6 & 134.9 &  554.6 & 994.2  \\
%\hline
&&&&&&&&  \\
\hline
 experiment & 139.57 & 493.68 & 497.61 & 92.1 & 110.1 & 134.98 &  547.86 & 957.78  \\
&&&& $\pm$0.8  &  $\pm$0.2  &&&  \\

\hline
\end{tabular}
\end{table}
%
% Eksperimentalne vrijednosti za fpi i fK su iz
% Leptonic decays of charged pseudoscalar mesons (rev.). Eq. (84.14)
% M. Tanabashi et al. (Particle Data Group),
% Phys. Rev. D 98, 030001 (2018) and 2019 update.

In the same way, the third row in every table starts with
the superscript label ${}^{(1)}$, because it results from
the fit aiming at the masses in Eq. (\ref{EMsubtractedM}).
It gives the FLAG \cite{Aoki:2013ldr} values for the $\pi^\pm$,
$K^\pm$ and $K^0$ just-QCD masses ({\it i.e.}, with the
corresponding electromagnetic contributions removed).
This fit yields $m_u = 3.40$ MeV, $m_d = 6.80$ MeV $= 2 m_u$
and $m_s = 115.61$ MeV, but the minimization procedure
did not reproduce the values in Eq. (\ref{EMsubtractedM})
exactly. Still, the difference between $M_{\pi^\pm}^{\rm fit}$,
$M_{K^\pm}^{\rm fit}$, $M_{K^0}^{\rm fit}$ in the row ${}^{(1)}$
and Eq. (\ref{EMsubtractedM}) -- stemming overwhelmingly from
 the kaon sector -- is satisfactorily small considering the
 rather large difference between the model quark mass parameter
 $m_s \approx 115$ MeV and the QCD $s$-quark current mass parameter
 $\mbox{m}_s = 93^{+11}_{-5}$ MeV \cite{Tanabashi:2018oca}.

For the three just described fittings, the first block of
 Table \ref{3M+2f} gives the corresponding versions of
 the three pion and kaon masses
which are protected by charge and/or strangeness from any
influence of the $U_A(1)$ anomaly even out of the isospin
symmetry limit, and to which we fitted the three parameters $m_q$.

The last block of Table \ref{3M+2f} gives, for our three fits,
 the predictions of our chosen DS  separable model for the 
 observable masses of the flavorless light pseudoscalar mesons 
 $\pi^0$, $\eta$ and $\eta'$. Out of the isospin limit,
 the neutral pion is not protected from the $U_A(1)$ anomaly,
 but its contribution, as well as the related admixture of
 the $s\bar s$ pseudoscalar bound state to $\pi^0$, is 
 of course small, since the isospin symmetry is very close
 to reality.

 The middle block of Table \ref{3M+2f} also contains our
 prediction of observables for the above fits,
 namely the decay constants of $\pi^+$ and $K^+$.
 
 Table \ref{auxiliary} gives mostly the quantities which,
 except the pion decay constant $f_{\pi}$, are not strictly 
observable. They are nevertheless presented, since they are
illustrative for the calculations outlined in the text above,
 especially of ${\bf M}^2_{NA}$,
 the non-anomalous part of the mass matrix.
% Hence, they are presented for completeness.

Table \ref{mqCondChi} gives quantities which enter
 into the calculation of ${\bf M}^2_{A}$,  the
 anomalous part of the mass matrix, since the products
 $m_q \, \langle \bar{q}q\rangle$ $(q=u,d,s)$ determine the
 QCD topological susceptibility $\chi$  (\ref{chiShore_small_m})
 and topological charge parameter $A$ (\ref{defA}).
 Their behavior at $T > 0$ will determine the fate of $U_A(1)$ 
 symmetry breaking and restoration in the future investigations
 in the present model. 

\begin{table}
\caption{For our three fits, the results for unphysical pseudoscalar bound
 states $|u\bar{u}\rangle$, $|d\bar{d}\rangle$, and $|s\bar{s}\rangle$: their
masses and decay constants (and also $f_{\pi^0} = (f_{u\bar u} + f_{d\bar d})/2$),
and the $SU(3)$ flavor and isospin symmetry breaking parameters
% defined from them by the respective ratios
$X\equiv f_{u\bar{u}}/f_{s\bar{s}}$ and $Y\equiv f_{u\bar{u}}/f_{d\bar{d}}$.
 All values are in MeV.}
\label{auxiliary}       % Give a unique label
\begin{tabular}{|c|ccccccccc|}
\hline
  &&&&&&&&& \\
Type of fit & $M_{u\bar{u}}$ & $M_{d\bar{d}}$ & $M_{s\bar{s}}$ & $f_{u\bar{u}}$ & $f_{d\bar{d}}$ & $f_{s\bar{s}}$ & $f_{\pi^0}$ & $X$ & $Y$ \\
  &&&&&&&&& \\
\hline
   &&&&&&&&& \\
$m_u=m_d$ & 140.1 & 140.0 & 685.0 & 92.0 &  92.0 & 119.0 & 92.0 & 0.773 & 1.0  \\
   &&&&&&&&& \\
${}^{(2)} m_u = 0.67\, m_d$ & 124.8 & 153.0 & 684.9 & 91.5 &  92.4 & 118.7 & 92.0  &  0.771 &  0.991  \\
   &&&&&&&&& \\
${}^{(1)} m_u = 0.50\, m_d$ & 110.0 & 155.9 & 684.9 & 91.1 &  92.5 & 118.7 & 91.8 & 0.768 &  0.986  \\
   &&&&&&&&&  \\
\hline
\end{tabular}
\end{table}

\begin{table}
\caption{For the old isosymmetric fit and the new fits, ${}^{(2)}$ and ${}^{(1)}$, with
broken isospin symmetry, $m_u \neq m_d$, the three sets of values of the model quark mass
parameters $m_q$ ($q=u,d,s$), are related to the model results for topological susceptibility
$\chi$ and ``massive'', {\it i.e.}, flavor-nonuniversal condensates $\langle\bar{u}u\rangle$,
$\langle \bar{d}d\rangle$ and $\langle \bar{s}s\rangle$.
Our model predictions for the topological susceptibility $\chi$ are evaluated
from these condensates and the corresponding $m_q$'s. These same sets of $m_q$'s
yield the corresponding values of the topological susceptibility $\chi_0$ when
 one uses the flavor-universal or ``massless'', {\it i.e.}, chiral-limit
 condensate $\langle \bar{q}q\rangle_0 = -217^3$ MeV${}^3$.
 All values are in MeV (or indicated powers of MeV).}
\label{mqCondChi}       % Give a unique label
\begin{tabular}{|c|cccccccc|}
\hline
        &&&&&&&&  \\
Type of fit  &      $m_u$ & $m_d$ & $m_s$ &$\chi_0$ & $\langle \bar{u}u\rangle$ & $\langle \bar{d}d\rangle$ & $\langle \bar{s}s\rangle$ & $\chi$  \\
        &&&&&&&&  \\
\hline
        &&&&&&&&  \\
$m_u=m_d$ & 5.49 &  5.49 & 115.12 & $72.18^4$ &  $-219^3$ &  $-219^3$ &  $-239^3$ &   $72.73^4$ \\
        &&&&&&&&  \\
${}^{(2)} m_u = 0.67\, m_d $ & 4.37 &  6.55 & 115.34 & $71.56^4$ &  $-219^3$ &  $-220^3$ &  $-239^3$ &   $72.22^4$ \\
 &&&&&&&&  \\
${}^{(1)} m_u = 0.50\, m_d $ & 3.40 &  6.80 & 115.61 & $69.09^4$ &  $-219^3$ &  $-220^3$ &  $-239^3$ &   $69.61^4$ \\
 &&&&&&&&  \\
\hline
\end{tabular}
\end{table}

\section{Summary}
\label{summary}

A well-tried DS effective model intended for investigations at $T > 0$
has been refitted at $T = 0$ by allowing its quark mass parameters
$m_q$ to take values out of the isospin limit. Potentially the
most significant improvement of the ensuing model parametrization
is lowering the quark mass parameter of the lightest flavor to
$m_u = m_d/2 = 3.40$ MeV, as explained in the rest of the text.

Namely, the isospin symmetry is mostly a very accurate 
approximation to reality, so that (as already hinted in
the Introduction), our aim of relaxing the isospin limit
is not a better description of the masses, decay constants
and other observables at $T = 0$. (But it is good to check
just in case, that nothing is spoiled by such refitting.)
Indeed, relaxing the isospin limit did not bring significant
 changes in directly observable quantities at $T = 0$.
 For example, even beyond the precision displayed in our
Tables, the calculated values of the two pion decay constants
remained unique, $f_{\pi^+} = f_{\pi^0}$, even for the fit 
yielding the larger difference between the two lightest flavors.
% $m_u = 3.40$ MeV $= m_d/2$.
 Some marginal improvement is seen
in the two last columns of Table \ref{3M+2f}, {\it i.e.}, the
masses of the isoscalar mesons $\eta$ and $\eta'$. The largest
improvement in predicting observable masses is the mass of the
neutral pion, but just relaxing the isosymmetric requirement
$m_u = m_d$ is of course not sufficient for that. This is
illustrated by the difference between our two fits beyond
the isosymmetric limit. Fitting cavalierly the empirical
masses of the charged pseudoscalar mesons imposes unjustifiably
their electromagnetic contributions on the predicted neutral
ones. If a model interaction mimics only the QCD one, fully
consistent out-of-isospin-limit fits should be to masses
 from which the electromagnetic contributions
have been taken out, such as Eq. (\ref{EMsubtractedM}).

Before explaining the potential importance of lowering the quark
mass parameters of the model, let us remark that we are of course
aware that as parameters of a phenomenological model, our $m_q$'s
cannot be related quite unambiguously and precisely to the still
somewhat lower values of the fundamental QCD current quark masses
 $\mbox{m}_u=2.16^{+0.49}_{-0.26}$ MeV,
 $\mbox{m}_d=4.67^{+0.48}_{-0.17}$ MeV and 
$\mbox{m}_s=93^{+11}_{-5}$ MeV \cite{Tanabashi:2018oca}.
The relationship of ratios is better defined, since
differences of various schemes tend to cancel in them. Hence,
our second ratio $m_u/m_d = 0.5$, within errors of
 the ratio of the lightest QCD current masses,
$\mbox{m}_u/\mbox{m}_d=0.47^{+0.06}_{-0.07}$ \cite{Tanabashi:2018oca},
 provides a better-defined connection between our
 model parameters and current quark masses of QCD.

Now, since the QCD topological susceptibility $\chi$ (\ref{chiShore_small_m})
 and topological charge parameter $A$ (\ref{defA}) depend on the products
 of quark masses and the corresponding condensates, 
$m_q\, \langle {\bar q}q\rangle$ ($q=u,d,s$), as their harmonic averages,
 the lightest flavor is dominant. While the absolute values of
 condensates fall rather slowly with the mass of their corresponding
 quark flavor towards the saturation at their limiting, chiral-symmetric
 value $\langle \bar{q}q\rangle_0$, their temperature dependence is
 a different story.  The crossover fall of a condensate with $T$
 quickly gets increasingly steeper for smaller values of $m_q$. So,
the steepest falling condensate is multiplied by the smallest quark mass.
 In conjunction with the harmonic average type of dependence in
 Eqs. (\ref{defA}) and (\ref{chiShore_small_m}),
 it is likely that reducing $m_u$ below its isospin partner $m_u$
 will significantly influence, {\it e.g.}, $T$-dependence of the
 $\eta'$-$\eta$ complex studied in Ref. \cite{Horvatic:2018ztu},
 and analogously various other cases of $U_A(1)$ restoration.
 This is why it is important to have models capable of
 investigating such situations also beyond the isospin limit.

\vspace{5mm}

\section*{Acknowledgment}
D. Klabu\v{c}ar thanks for partial support to  
the organizers of 40. Max Born Symposium
``Three Days on Strong Correlations in Dense Matter'',
and to COST Actions CA15213 THOR and CA16214 PHAROS.

\end{document}